\documentclass[aps, prl, twocolumn, superscriptaddress,preprintnumbers]{revtex4-2}
\usepackage{bm, amsmath, amsfonts, amssymb, ascmac, mathtools, braket}
\usepackage{times}
\usepackage{multirow}
\usepackage{graphicx}
\usepackage{float, color, xcolor}

\usepackage[whole]{bxcjkjatype} 
\usepackage{subfigure} 
\usepackage{amscd} 

\usepackage{bbm}
\usepackage{tabularx}

\usepackage{comment}

\usepackage[
pagebackref=false,
colorlinks=true,
linkcolor=blue,
urlcolor=blue,
filecolor=black,
citecolor=red,
pdfstartview=FitV,
pdftitle={},
pdfauthor={},
pdfsubject={},
pdfkeywords={},
pdfpagemode=None,
bookmarksopen=true
]{hyperref}

\newcommand{\ii}{\text{i}}

\begin{document}

\title{Callan-Rubakov effects in topological insulators}

\author{Yusuke O. Nakai}
\affiliation{Center for Gravitational Physics and Quantum Information, Yukawa Institute for Theoretical Physics, Kyoto University, Kyoto 606-8502, Japan
}
\affiliation{
Max-Planck-Institut f\"ur Festk\"orperforschung, Heisenbergstrasse 1, D-70569 Stuttgart, Germany
}

\author{Reuel Dsouza}
\affiliation{
Max-Planck-Institut f\"ur Festk\"orperforschung, Heisenbergstrasse 1, D-70569 Stuttgart, Germany
}

\author{Daichi Nakamura}
\affiliation{Institute for Solid State Physics, University of Tokyo, Kashiwa, Chiba 277-8581, Japan}

\author{Shu Hamanaka}
\affiliation{Department of Physics, Kyoto University, Kyoto 606-8502, Japan}
\affiliation{Institute for Theoretical Physics, ETH Zurich, 8093 Zurich, Switzerland}

\author{Andreas P. Schnyder}
\affiliation{
Max-Planck-Institut f\"ur Festk\"orperforschung, Heisenbergstrasse 1, D-70569 Stuttgart, Germany
}

\author{Masatoshi Sato}
\affiliation{Center for Gravitational Physics and Quantum Information, Yukawa Institute for Theoretical Physics, Kyoto University, Kyoto 606-8502, Japan
}

\date{\today}

\begin{abstract}
The Callan-Rubakov effect describes
monopole-catalyzed proton decay. 
While this effect is fundamental for quantum field theories, its experimental observation has remained far from reality. 
Here, we reveal a similar, but experimentally reachable, defect-catalysis of the quantum anomaly in topological materials. 
In particular, surface Dirac fermions on topological insulators develop a distinct localized state at the position of dislocations or $\pi$-fluxes, which mediates spin-flip time-reversal breaking scattering or absorption of electrons.
Despite the Hermiticity of topological insulators, a non-Hermitian topological number guarantees the robust existence of the localized state. 
Our finding implies that non-magnetic defects may behave like magnetic impurities on surfaces of topological insulators.
Using  the K-theory classification, we 
generalize this condensed-matter version of
the Callan-Rubakov effect to other classes of topological materials.

\end{abstract}

\maketitle

{\it Introduction.}---
One of the most intriguing aspects of topological materials is their ability to realize various theoretical ideas of particle physics in condensed matter systems.    
Their topological excitations mimic relativistic fermions, such as Dirac fermions, enabling their intrinsic phenomena in non-relativistic electron systems.  
For instance, a Chern insulator hosts a one-dimensional chiral fermion as a topological edge state, which exhibits chiral anomaly in the quantum Hall effect \cite{q1-quantum-hall-experiment-Klitzing,q2-Laughlin-conductivity,q3-TKNN,q4-Thouless-transport,q5-Kohmoto-1985,q6-Haldane-1998,witten2016three}.
Dirac-Weyl semimetals unveil the generation of an electric current induced by a chirality imbalance in the presence of a magnetic field, the chiral magnetic effects \cite{cm1-fukushima2008,li2016chiral,son2013,burkov2014,zhang2016signatures,
cm7-Vozmediano-2016,cm8-franz-2016,cm4-shumiyoshi-2016,
cm3-taguchi-2016, cm6-higashikawa-2019,cm9-Budich-2019,
NN-bessho-sato-2021}. 
A monopole configuration in topological insulators acquires a fractional charge due to the Witten effect~\cite{Witten-PL-79,TQFT-2008-TRS,franz-2010}. 
Furthermore, the self-antiparticle Majorana fermions can emerge in topological superconductors or topological spin-liquids, displaying the non-Abelian anyon statistics \cite{read-green2000,ivanov2001, kitaev2001unpaired, sato2003non,fu-kane2008,sato2009,lutchyn2010,oreg2010, alicea2011non,kitaev2006anyons,jackeli2009}. Also, more exotic space-time supersymmetry may emerge at the boundary of a topological phase \cite{SUSY-Vishwanath,susy-Affleck-2015,yao2017}. 

While many interesting ideas about particle physics have been discussed in topological phases, the application of defect-catalysis of anomalous processes to condensed matter physics has not yet been explored.
In particular, the Callan-Rubakov effect~\cite{Rubakov-NP-82, Callan-PRD-82},
describing monopole-catalyzed proton decay,
is worth to be addressed. 
Whereas the Callan-Rubakov effect is fundamental for quantum field theories, experimental observation is far from reality because both the monopole and proton decay have yet to be discovered. 

In this paper, we pursue analog defect-mediated effects for quantum anomalies in topological insulators and superconductors. 
In particular, we discover that a surface Dirac fermion on a three-dimensional time-reversal invariant topological insulator develops a distinct localized state at the intersection between a line defect and a surface.
The localized state results in the following observable effects:
(i) Enhanced local density of states, (ii) absorption of an electron, and (iii)  
spin-flip time-reversal breaking scattering, due to the quantum anomaly.
\begin{figure}[tp]
 \begin{center}
  \includegraphics[scale=0.09]{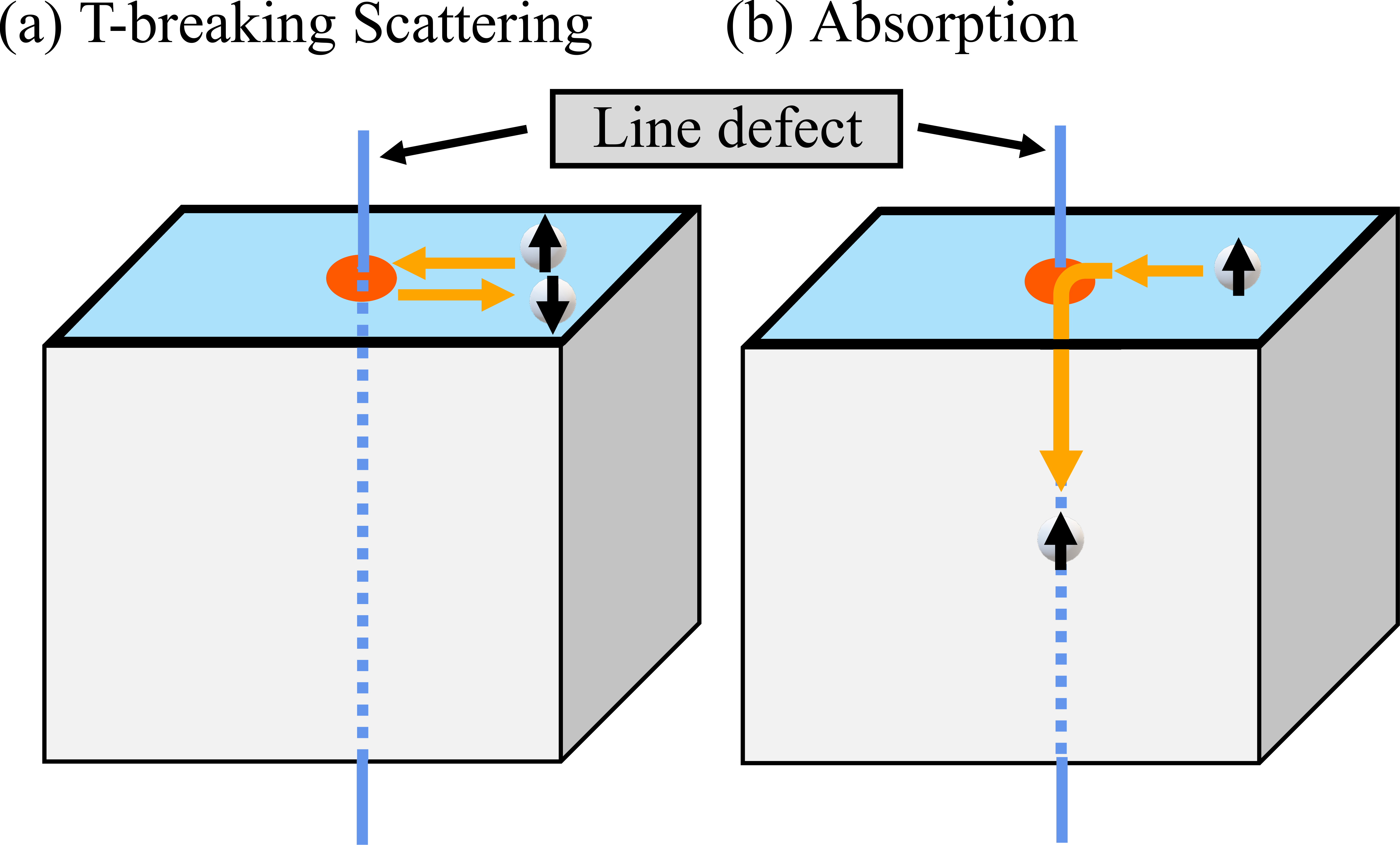}
  \caption{Two possible processes in the Callan-Rubakov effect in topological insulators. (a) Spin-flip time-reversal breaking scattering. 
  The time-reversal anomaly induces a magnetic condensate mediating the time-reversal breaking scattering at the intersection between the surface and the line defect. (b) Electron absorption. An absorbed electron becomes a one-dimensional gapless mode along the line defect.
 \label{two possibility}}
  \end{center}
\end{figure}
In particular, the last effect implies that the non-magnetic line defect may behave like a magnetic impurity at the intersection with the surface.

\begin{figure}[btp]
 \begin{center}
  \includegraphics[scale=0.08]{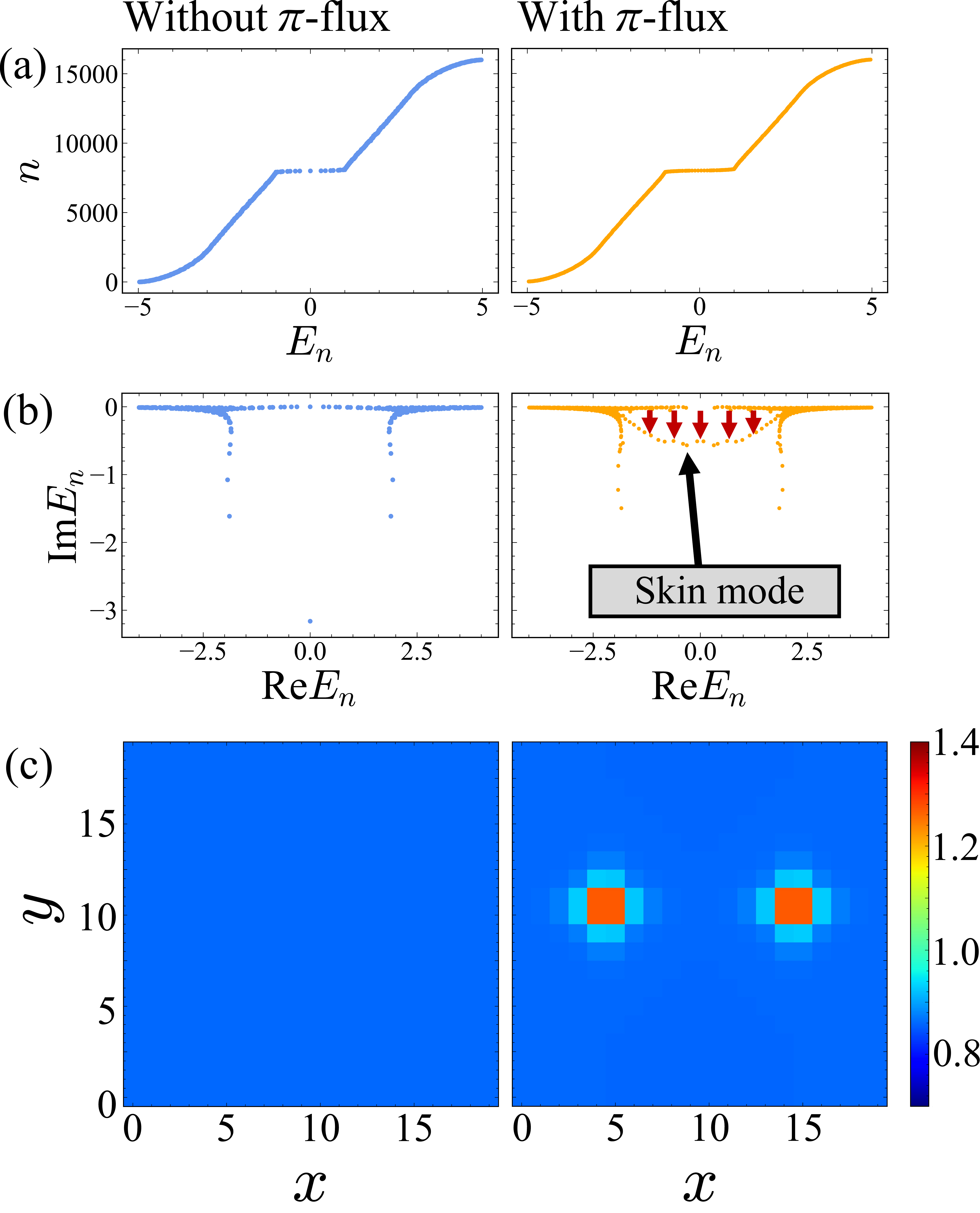}
  \caption{Modes relevant to the Callan-Rubakov effect with $\pi$-flux. Periodic boundary conditions are imposed along the $x$ and $y$ directions, while open boundary conditions are applied along the $z$ direction ($L_x=L_y=20, L_z=10, t=1.0, m=-2.0$).  The $\mathbb{Z}_2$ indices are $\nu_0=\pi$, ${\bm M}_\nu=(0,0,0)$. (a) Energy spectra of the total Hermitian Hamiltonian without $\pi$-flux (left) and with $\pi$-flux (right). (b) Complex spectra of the effective surface non-Hermitian Hamiltonian $H_{\rm eff}$ without $\pi$-flux (left) and with $\pi$-flux (right). Every eigenvalue is two-fold degenerate due to the time-reversal symmetry.  
  (c) Local density of states at the surface without $\pi$-flux (left) and with $\pi$-flux (right).
    The infinitesimal number $\eta$ in the retarded Green's function is chosen as $\eta=1/\sqrt{L_z}$ \cite{hamanaka2024nonhermitiantopologyhermitiantopological}.
  \label{fig_flux_hermitian_non-Hermitian_eigenstate_LDOS}}
 \end{center}
\end{figure}

To reveal the robust existence of the localized state, we employ a recently proposed formulation for topological boundary states \cite{schindler-BBC-2023,
inaka-point,
hamanaka2024nonhermitiantopologyhermitiantopological}. 
A well-defined localized state in topological phases requires a gap from other states 
\cite{avron1983,
Hasan-review-2010,TS1-review-2011,Chingkai-teo-andreas-ryu}.
For instance, a surface gap of higher-order topological insulators is essential for identifying a topological intersection state reported recently \cite{schindler2022defectNature}.  
However, such a conventional identification does not work for the present intersection states because the surface of a (first-order) topological insulator is already gapless (Fig. \ref{fig_flux_hermitian_non-Hermitian_eigenstate_LDOS} (a)).
In contrast, our formulation of the boundary states enables the identification of localized states in the gapless boundaries. 
Regarding the bulk as an environment of the boundary, the new formulation describes the boundary in terms of a non-Hermitian effective Hamiltonian.
The localized state is well-separated from other gapless states in the complex energy plane of the effective Hamiltonian
(Fig. \ref{fig_flux_hermitian_non-Hermitian_eigenstate_LDOS} (b)).
We also show that a non-Hermitian topological number ensures the robust existence of the localized state.

Our theory also applies to other topological materials.
Employing the K-theory classification, we 
generalize the topological-insulator-based
Callan-Rubakov effect to other topological materials with line defects.

{\it Surface Dirac fermions coupled to $\pi$-flux}---
To illustrate the Callan-Rubakov effect in topological insulators, let us start with a two-dimensional Dirac fermion on the surface of a three-dimensional time-reversal invariant topological insulator.
The Hamiltonian of the surface Dirac fermion is
$
H= \psi^\dagger{\cal H}\psi$, with ${\cal H}=-i\partial_xs_x-i\partial_ys_y$,
where $\psi$ is a two-component spinor, $(x,y)$ is the coordinate of the surface, and $s_{i=x,y}$ is the Pauli matrix for spin.
The Hamiltonian exhibits time-reversal symmetry, $T{\cal H}T^{-1}={\cal H}$, with $T=is_y K$, where $K$ is the complex conjugation operator.
Time-reversal symmetry prohibits a mass term of the surface Dirac fermion.

To couple the system to a $\pi$-flux, we use the circular coordinate $(x,y)=(r\cos\theta, r\sin\theta)$, which leads to
\begin{align}
{\cal H}=
\begin{pmatrix}
0 & e^{-i\theta}(-i\partial_r-\partial_\theta/r)\\
e^{i\theta}(-i\partial_r+\partial_\theta/r) & 0
\end{pmatrix}.
\label{eq: H}
\end{align}
Because of rotation symmetry, the total spin normal to the surface, $J_z=-i\partial_\theta+s_z/2$, is a good quantum number for $\psi$, $J_z\psi=m_z\psi$.
Then, we introduce a $\pi$-flux as the anti-periodic condition of $\psi$ in the circular direction $\theta$, $\psi(\theta+2\pi)=-\psi(\theta)$. The coupling to the $\pi$-flux converts the spin $m_z$ of the Dirac fermion into an integer.

Now, let us focus on the $J_z=0$ sector, $\psi_{J_z=0}=\psi_++\psi_-$ with $\psi_\pm=
(e^{-i\theta/2}, \pm e^{i\theta/2})^t\chi_\pm /\sqrt{4\pi r}
$. 
The Dirac equation with Eq. (\ref{eq: H}) leads to $i\partial_t\chi_{\pm}=\mp i\partial_r\chi_\pm$, whose solution is
\begin{align}
\chi_\pm=e^{-iE(t\mp r)}.    
\label{eq: chi}
\end{align}
Here we note that $\psi_{\pm}$ have special properties: (i)
$\psi_{+}$ 
is purely an outgoing wave from the $\pi$-flux, while $\psi_-$ is purely an incoming wave.
(ii) They form a Kramers pair, $\psi_-=is_y K\psi_+$.
(iii)
Both $\psi_\pm$ are singular at the origin, the location of the $\pi$-flux, so there is no obvious criterion for the boundary condition to match the incoming and outgoing waves at the origin. 

The peculiarity of these properties becomes evident when considering how the $\pi$-flux scatters the $J_z=0$ incoming electron $\psi_-$.
Considering the preservation of $J_z$, we have only two possibilities: 
The $\pi$-flux scatters the incoming electron into the $J_z=0$ outgoing electron $\psi_+$ or absorbs the incoming electron (Fig. \ref{two possibility}). While the former possibility might look reasonable, it requires time-reversal breaking. 
Otherwise, the incoming wave $\psi_-$ cannot be scattered into its Kramers partner $\psi_+$ because of the Kramers theorem.
In other words, the former suggests that the $\pi$-flux can induce a magnetic order that mediates the time-reversal breaking scattering for surface Dirac fermions on topological insulators.  
As described below, we find that both processes may occur.  

Notably, the above effect is closely analogous to the Callan-Rubakov effect.
The original Callan-Rubakov effect treats three-dimensional chiral fermions coupled to a magnetic monopole and examines the $J=0$ sector using rotation symmetry \cite{Rubakov-NP-82, Callan-PRD-82}.
Then, the left-handed (right-handed) chiral fermion with charge +1 (-1) is found to be a purely incoming wave, while the right-handed (left-handed) one with charge +1 (-1) is a strictly outgoing wave \cite{Rubakov-review-88,
supplement}.
Therefore, once the chiral fermion reaches the monopole, it must convert its chirality or charge after the scattering. 
These unusual scattering processes catalyze the proton decay in the grand unified theory, 
and are called the Callan-Rubakov effect.
From this similarity, we dub the above anomalous scattering the Callan-Rubakov effect in topological insulators. 

{\it The Callan-Rubakov effect in topological insulators.}---
We next describe how the electron absorption and the spin-flip scattering occur at the $\pi$-flux.
We first discuss the absorption (Fig. \ref{two possibility} (b)).
The absorption process becomes evident when returning to the full three-dimensional description of a topological insulator. 
As illustrated in Fig. \ref{two possibility},
the $\pi$-flux goes through the bulk topological insulator. 
Remarkably,
a one-dimensional gapless helical mode exists along the $\pi$-flux in the present case due to the bulk-defect correspondence \cite{teo-kane-defect}.
Therefore,
when the incoming surface electron hits the $\pi$-flux, it can flow into the bulk along the $\pi$-flux, and thus the absorption occurs.

On the other hand, for the spin-flip scattering process (Fig.~\ref{two possibility} (a)), we need to take into account the quantum anomaly.
As mentioned above, the scattering process requires time-reversal breaking.
As the chirality change occurs in the original Callan-Rubakov effect due to the chiral anomaly \cite{Rubakov-review-88}, a similar quantum anomaly is essential for the spin-flip scattering process. 
As shown in the following, for the surface Dirac fermion of a topological insulator, time-reversal symmetry 
is incompatible with the electromagnetic gauge symmetry due to the quantum anomaly, and thus the required time-reversal breaking occurs at the $\pi$-flux.   

To see the incompatibility, we couple the massless Dirac fermion in Eq.(\ref{eq: H}) with the electromagnetic gauge fluctuations $A_\mu(x,y,t)$ $(\mu=x,y,t)$ on the surface. 
The total Lagrangian reads 
\begin{align}
L=-\frac{1}{4e^2}\int dx^2 F_{\mu\nu}F^{\mu\nu}+
\int dx^2 i\bar{\psi}\Gamma^\mu D_\mu\psi,    
\end{align}
where $\bar{\psi}=\psi^\dagger\Gamma^t$, $D_\mu=\partial_\mu-iA_\mu$, $\Gamma^t=s_z$, $\Gamma^x=-is_y$, and $\Gamma^y=is_x$.
In the leading order, the $J_z=0$ electron, $\psi_{J_z=0}=\psi_+ + \psi_-$, couples to the $s$-wave fluctuation $A_\mu=(a_t(r,t), a_r(r,t)\cos\theta, a_r(r,t)\sin\theta)$. 
Substituting them for the Lagrangian, we have the dominant contribution as a variant of the massless Schwinger model
\begin{align}
L_{J_z=0}=\int dr\left[-\frac{\pi r}{2e^2}f_{\alpha\beta}f^{\alpha\beta}+i\bar{\chi}\gamma^\alpha
(\partial_\alpha-ia_\alpha)\chi\right],   
\label{eq:Schwinger}
\end{align}
where $\alpha,\beta=t,r$, $f_{\alpha\beta}=\partial_\alpha a_\beta-\partial_\beta a_\alpha$,  
$\chi=(\chi_+,\chi_-)^t$, $\bar{\chi}=\chi^\dagger\gamma^t$, $\gamma^t=\tau_x$, and $\gamma^r=-i\tau_y$ ($\tau_i$ is the Pauli matrix acting on the $\chi$ grading).
Note that correlation effects become prominent at the location of the $\pi$-flux, because of the $r$-dependent electromagnetic coupling $e^2_{\rm eff}(r)=e^2/2\pi r$. 
Using bosonization, $L_{J_z=0}$ is recast into \cite{supplement}
\begin{align}
L_{J_z=0}=\int dr \left[\frac{1}{e^2_{\rm eff}(r)}f_{tr}^2+\frac{1}{2\pi}f_{tr}\phi+\frac{1}{8\pi}\partial_\alpha\phi\partial^\alpha\phi
\right],     
\end{align}
which leads to the mass term of the boson $\phi$ as $(e^2_{\rm eff}(r)/8\pi^2)\phi^2$ after integrating out the gauge fluctuation $f_{tr}$.
Since the boson's mass diverges at the origin, $\phi$ goes to zero at $r\rightarrow 0$, which implies the finite condensation $\langle \bar{\chi}\chi\rangle=\langle \chi^\dagger_+\chi_-+\chi^\dagger_-\chi_+\rangle
\sim \cos\phi\neq 0$ at $r=0$.
The condensation mixes $\chi_+$ and $\chi_-$, providing the required time-reversal breaking term.
In other words, the $\pi$-flux can behave like a magnetic impurity on surfaces of topological insulators, 
enabling the spin-flip scattering process (Fig. \ref{two possibility} (a)).

Note that the states in Eq.~(\ref{eq: chi}) enhance the local density of states at the defect:
These states have a momentum-independent density of states,
since they have linear energy dispersion for the radial momentum direction. 
Therefore, their amplitude squared $|\psi_{\pm}|^2\sim 1/r$ gives their local density of states, which diverges at the $\pi$-flux \cite{supplement}.
The enhanced density of states mediates the scattering or absorbing processes mentioned above.

{\it The Callan-Rubakov effect as the non-Hermitian skin effect.}---
Like the original Callan-Rubakov effect, our argument relies on rotation symmetry so far:
We specify a mode responsible for the Callan-Rubakov effect from the surface Dirac fermion using the eigenvalue of $J_z$. 
However, actual topological insulators do not have such rotation symmetry. 
Now, we identify the Callan-Rubakov effect without relying on rotation symmetry. 

A crucial observation is the uni-directed motion of the $J_z=0$ Dirac fermion in the Callan-Rubakov effect:
$\psi_+$ ($\psi_-$) is a purely outgoing (incoming)  wave.
Notably, such a uni-directed motion is typical for one-dimensional topological modes.
For instance, topological helical edge modes in quantum spin Hall states exhibit similar uni-directed helical motions \cite{kane-mele2005,bernevig2006}.
Therefore, the mode relevant to the Callan-Rubakov effect should have a topological origin.
However, 
a conventional topological number for a gapped system is useless for characterizing the relevant mode, since the surface Dirac fermion is already gapless (Fig.~\ref{fig_flux_hermitian_non-Hermitian_eigenstate_LDOS} (a)).  
In other words, we must somehow extract the one-dimensional gapless topological mode $\psi_{\pm}$ from the two-dimensional gapless Dirac fermion.

We overcome this difficulty by using a new description of boundary states \cite{inaka-point,hamanaka2024nonhermitiantopologyhermitiantopological}.
A distinct feature of the topological boundary states is the accompanying bulk states:
They are indispensable to compensate for the quantum anomaly of the surface states.
Thus, we can better describe the surface state by considering an influence from the bulk.  
Notably, surface states now may have a finite lifetime,  {\it i.e.} their energy may acquire an imaginary part, due to the coupling to the bulk as an environment. 
As a result, an effective non-Hermitian Hamiltonian $H_{\rm eff}$ describes the surface states, although the original Hamiltonian $H$ is Hermitian. 
As shown in \cite{supplement}, one can construct such $H_{\rm eff}$ systematically using the retarded Green's function.

Once we adopt the non-Hermitian description of the surface state, 
we can identify the relevant one-dimensional gapless mode among others in the complex energy plane (Fig. \ref{fig_flux_hermitian_non-Hermitian_eigenstate_LDOS} (b)):
Even though all the surface modes are gapless in the real part of the energy, the relevant mode is distinct in the complex energy plane of $H_{\rm eff}$.
Furthermore, we can identify the relevant mode as a topological mode in $H_{\rm eff}$: 
From non-Hermiticity, the characteristic uni-directed motion results in a macroscopic accumulation of states at the $\pi$-flux, known as a ($\pi$-flux) non-Hermitian skin effect~\cite{lee-skin-2016,yao-skin-2018, r8-kunst-biorthogonal-2018, Gong-ashida-2018,r18-lee-2019,r9-yokomizo-murakami-nonbloch-2019,
Robert,
okss, Zhang-2020, dislocation-Fulga2021, r21-okuma-anomaly-2021,Schindler-dislocation2021, 
Sun-2021,Panigrahi-2022,
r6.2okuma-sato-review, Schindler-magnetic2023, 
Chong-Disclination2024}.
The skin effect originates from a topological number intrinsic to non-Hermitian systems:
From the correspondence between the Hermitian bulk topology in spatial dimensions $d$ and the non-Hermitian boundary topology in spatial dimensions $d_{\rm eff}=d-1$~\cite{Lee-PRL-19, NN-bessho-sato-2021, inaka-point, hamanaka2024nonhermitiantopologyhermitiantopological}, the Hermitian $\mathbb{Z}_2$ topology of the three-dimensional ($d=3$) topological insulator results in the two-dimensional ($d_{\rm eff}=2$) non-Hermitian $\mathbb{Z}_2$ topology of the surface effective Hamiltonian.
Then, from an argument by Teo and Kane~\cite{teo-kane-defect}, the $\pi$-flux with codimension $D=1$, decreases the dimension of the non-Hermitian
$\mathbb{Z}_2$ number by 1, {\it i.e.} $\delta_{\rm eff}\equiv d_{\rm eff}-D=2-1=1$. Hence, the surface with the $\pi$-flux hosts a one-dimensional ($\delta_{\rm eff}=1)$ non-Hermitian $\mathbb{Z}_2$ topology, which explains an accumulation of a one-dimensional mode as a consequence of the $\mathbb{Z}_2$ non-Hermitian skin effect \cite{okss}.
Therefore, we can identify the uni-directed mode for the Callan-Rubakov effect as the non-Hermitian skin mode. 

We confirm the validity of the above argument in a model of a three-dimensional topological insulator:
\begin{align}\label{eq:hamil-TI}
H({\bm k})&=t\sin k_x \sigma_zs_y-t\sin k_y\sigma_z s_x+t\sin k_z\sigma_y s_0
\nonumber\\
&+(m+t\cos k_x+t\cos k_y+t\cos k_z)\sigma_x s_0,
\end{align}
where $t$ and $m$ are real parameters, $\sigma_i$ and $s_i$ are the Pauli matrices in the orbital and spin spaces, respectively, $s_0$ is the $2\times 2$ identity matrix in the spin space, and $k_i$ are the momenta $(i=x,y,z)$.
As shown in Fig. \ref{fig_flux_hermitian_non-Hermitian_eigenstate_LDOS} (b), the complex spectrum of the surface Hamiltonian $H_{\rm eff}$ with the $\pi$-flux exhibits a well-separated one-dimensional mode in the complex energy plane, which is absent in the spectrum without the $\pi$-flux.
Moreover, the accumulated mode displays an enhanced local density of states, see Fig. \ref{fig_flux_hermitian_non-Hermitian_eigenstate_LDOS} (c). 
As expected from the analysis above for the Callan-Rubakov effect, 
the enhanced local density of states shows the power-law behavior $1/r$ as a function of the distance $r$ from the $\pi$-flux \cite{supplement}. 
All these results are consistent with our identification of the Callan-Rubakov effect as the non-Hermitian skin effect.

\begin{figure}[btp]
 \begin{center}
  \includegraphics[scale=0.09]{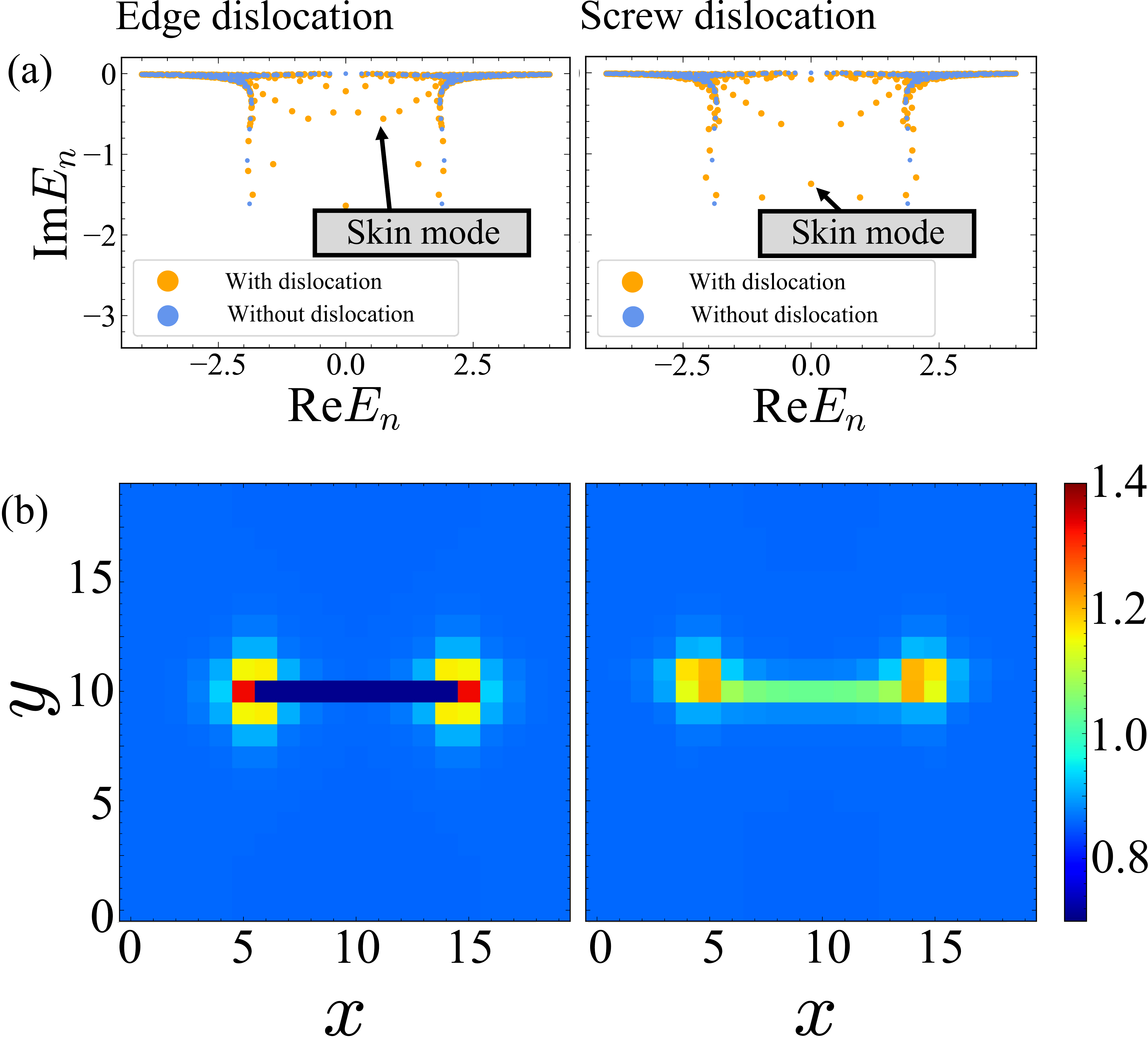}
  \caption{Modes relevant to the Callan-Rubakov effect with dislocations. We impose periodic boundary conditions in the $x$ and $y$-directions, and open boundary condition in the $z$-direction ($L_x=L_y=20, L_z=10, t=1.0, m=2.0$). The $\mathbb{Z}_2$ indices are $\nu_0=\pi$, ${\bm M}_\nu=(\pi,\pi,\pi)$.
  (a) (Left) Complex spectra of the effective surface non-Hermitian Hamiltonian $H_{\rm eff}$ with (orange) and without (blue) edge dislocation. The Burgers vector is ${\bm B}=(0,1,0)$. 
  (Right) Complex spectra of the effective surface non-Hermitian Hamiltonian $H_{\rm eff}$ with (orange) and without (blue) screw dislocation. The Burgers vector is ${\bm B}=(0,0,1)$. All eigenvalues exhibit Kramers degeneracy due to the time-reversal symmetry. (b) Local density of states (LDOS) at the surface of the topological insulator in Eq.~(\ref{eq:hamil-TI}) for each dislocation.}\label{fig_edge_screw_energy_eigenstate_LDOS}
  \end{center}
\end{figure}

The new theory enables us to pursue the Callan-Rubakov effect in other defects without rotation symmetry. 
Below, we reveal the Callan-Rubakov effect for general line defects in crystals, {\it i.e.} dislocations. 
Consider a three-dimensional topological insulator with the strong index $\nu_0$ and the weak indices ${\bm M}_\nu=(\nu_x,\nu_y,\nu_z)$, and insert a dislocation with the Burgers vector ${\bm B}$. 
Such a dislocation supports a one-dimensional helical mode when  ${\bm B}\cdot{\bm M}_{\nu}=\pi$ $({\rm mod.} 2\pi)$~\cite{ran2009dislocation}. 
Then, consider a surface normal to the dislocation. 
We find that the surface Hamiltonian under this situation exhibits the dislocation non-Hermitian skin effect \cite{dislocation-Fulga2021,Schindler-dislocation2021}, both for edge and screw dislocations, as shown in Fig. \ref{fig_edge_screw_energy_eigenstate_LDOS}. (See also \cite{supplement}.)
Therefore, the surface supports the uni-directed modes indispensable for the Callan-Rubakov effect.
For $\nu_0=\pi$, we also observe the enhancement of the local density of states at the intersection between the surface and the dislocation, which is necessary for the time-reversal breaking condensation (Fig. \ref{fig_edge_screw_energy_eigenstate_LDOS}).
Therefore, like the $\pi$-flux, we expect both the electron absorption and the time-reversal breaking spin-flip scattering at the dislocation.

\begin{table*}[t]
	\centering
	\caption{Topological table for the Callan-Rubakov effect in three-dimensional topological insulators/superconductors with line defects. The presence ($\pm 1$) and/or absence (0) of time-reversal symmetry (TRS), particle-hole symmetry (PHS), and chiral symmetry (CS) specify the ten-fold Altland-Zirnbauer (AZ) symmetry classes. The Hermitian topology in spatial dimension $d=3$ and the defect non-Hermitian topology in $\delta_{\rm eff}=1$ ensures gapless surface fermions and undirected modes to line defects, respectively, both of which are necessary for the Callan-Rubakiov effect in three-dimensional topological insulators/superconductors. (See classes DIII and AII.) } 
	\label{tab: classification}
     \begin{tabular}{c|ccc|cc} \hline \hline
    ~~AZ class~~ & ~~TRS ~~ & ~~PHS ~~ & ~~CS~~ & ~~$\delta=2/\delta_{\rm eff}=1$~~ & ~~$d=3/d_{\rm eff}=2$~~ \\ 
    \hline
    A & $0$ & $0$ & $0$ &  $\mathbb{Z}$ & $0$  \\
    AIII & $0$ & $0$ & $1$ &  $0$ & $\mathbb{Z}$ \\ \hline
    AI & $+1$ & $0$ & $0$ & $0$ & $0$ \\
    BDI & $+1$ & $+1$ & $1$  & $0$ & $0$  \\
    D & $0$ & $+1$ & $0$  & $\mathbb{Z}$  &0 \\
    DIII & $-1$ & $+1$ & $1$  & $\mathbb{Z}_2$ & $\mathbb{Z}$  \\
    AII & $-1$ & $0$ & $0$ & $\mathbb{Z}_2$ & $\mathbb{Z}_2$ \\
    CII & $-1$ & $-1$ & $1$  & $0$ & $\mathbb{Z}_2$\\
    C & $0$ & $-1$ & $0$ & $2\mathbb{Z}$  & $0$  \\
    CI & $+1$ & $-1$ & $1$  & $0$ & $2\mathbb{Z}$  \\ \hline \hline
  \end{tabular}
\end{table*}


{\it Generalization.}---
Our theory also paves the way for generalizing the Callan-Rubakov effect to other classes of topological materials.
As a straightforward generalization, we focus on three-dimensional topological materials with line defects for simplicity.
Since the generalized Callan-Rubakov effect needs gapless surface fermions and the defect skin effect, it requires two independent topological numbers: (i) A three-dimensional ($d=3$) Hermitian topological number of the original bulk Hamiltonian $H$, which ensures gapless surface fermions, 
and (ii) a one-dimensional ($\delta_{\rm eff}=1$) defect non-Hermitian topological number of the surface Hamiltonian $H_{\rm eff}$, which is indispensable for the defect skin effects.  
Table \ref{tab: classification} summarizes the $d=3$ bulk Hermitian topology and the $\delta_{\rm eff}=1$ defect non-Hermitian topology in the K-theory classification for the Altland-Zirnbauer (AZ) classes \cite{supplement}.
From the correspondence between the Hermitian topology in $H$ and the non-Hermitian one in $H_{\rm eff}$, the same table also gives the $d_{\rm eff}=2$ non-Hermitian topology and the $\delta=1$ defect Hermitian topology: A nontrivial $d=3$ bulk Hermitian topology ($\delta_{\rm eff}=1$ non-Hermitian defect topology) implies a nontrivial $d_{\rm eff}=2$ surface non-Hermitian topology ($\delta\equiv d-D=2$ defect Hermitian topology).  
Remarkably, line-defect skin effects on the surfaces always accompany defect gapless modes along the line defect, enabling fermion absorption as discussed above.

Table \ref{tab: classification} indicates that in addition to the three-dimensional time-reversal invariant topological insulators (class AII) considered above,  
three-dimensional time-reversal invariant topological superconductors, which belong to class DIII, may exhibit the line-defect Callan-Rubakov effect.
For class DIII systems, 
the surface supports an odd number of two-dimensional Majorana fermions when the bulk three-dimensional winding number is odd. 
The effective Hamiltonian is the same as Eq. (\ref{eq: H}), but $\psi$ is now a Majorana fermion. 
Similarly to topological insulators, the surface Majorana fermions show a non-Hermitian skin effect when a line defect penetrates the surface.
The enhanced local density of states appears at the interface between the defect and the surface, and the line defect hosts a one-dimensional helical Majorana mode; both are necessary for the scattering and the absorption 
in the Callan-Rubakov effect. 
Although the Majorana fermion has no direct electromagnetic coupling, 
it has an Ising-type magnetic dipole moment \cite{chung2009, shindou2010, mizushima2012}. 
Thus, the magnetic fluctuations due to electron-electron interactions could lead to the required magnetic order for the spin-flip scattering.

{\it Discussions.}--- In this Letter, we discuss novel line defect-mediated topological phenomena on surfaces of topological insulators. In particular, we reveal that the intersection between a line defect and a surface on a topological insulator behaves as an electron absorber or a magnetic impurity, mimicking the Callan-Rubakov effect. Whereas the system is intrinsically Hermitian, non-Hermitian topology guarantees the robustness of the effect. 

Our new characterization of the Callan-Rubakov effect applies to various systems. For instance, a monopole in a Weyl semimetal is expected to show a similar effect due to the monopole skin effect \cite{r21-okuma-anomaly-2021}. In addition, our theory could open a possible formulation of the original Callan-Rubakov effect in the lattice gauge theories. It is worthwhile to pursue such extensions.

\vspace{1ex}
{\it Acknowledgments.--}
We thank Kohei Kawabata and Keisuke Totsuka for their helpful discussions.
Y.O.N. is supported by JST SPRING, Grant Number JPMJSP2110, the JSPS Overseas Challenge Program for Young Researchers, and the MEXT WISE Program. Y.O.N. is grateful for the support and hospitality of the Max Planck Institute for Solid State Research. R.D. is supported by the IMPRS-CMS MSc fellowship program. D.N. is supported by JSPS KAKENHI Grant No.~24K22857. S.H. is supported by JSPS Research Fellow No.~24KJ1445, JSPS Overseas Challenge Program for Young Researchers, MEXT WISE Program, and the Pauli Center for Theoretical Studies.
A.P.S. is funded by the Deutsche Forschungsgemeinschaft (DFG, German Research Foundation) – TRR 360 – 492547816.
M.S. is supported by JSPS KAKENHI (Grants No. JP24K00569) and JST CREST (Grants No. JPMJCR19T2).
We thank the hospitality of the Simons Center for Geometry and Physics at Stony Brook University.

\bibliography{bib}

\widetext
\pagebreak
\begin{center}
{\bf \large Supplemental Material}
\end{center}

\renewcommand{\theequation}{S\arabic{equation}}

\renewcommand{\thefigure}{S\arabic{figure}}
\renewcommand{\thetable}{S\arabic{table}}
\renewcommand{\thesection}{S\arabic{section}}

\setcounter{equation}{0}
\setcounter{figure}{0}
\setcounter{table}{0}
\setcounter{section}{0}

\section{S1. Non-Hermitian topology in Hermitian topological matter}\label{sec:nH_in_H}

\begin{figure}[h]
 \begin{center}
  \includegraphics[scale=0.13]{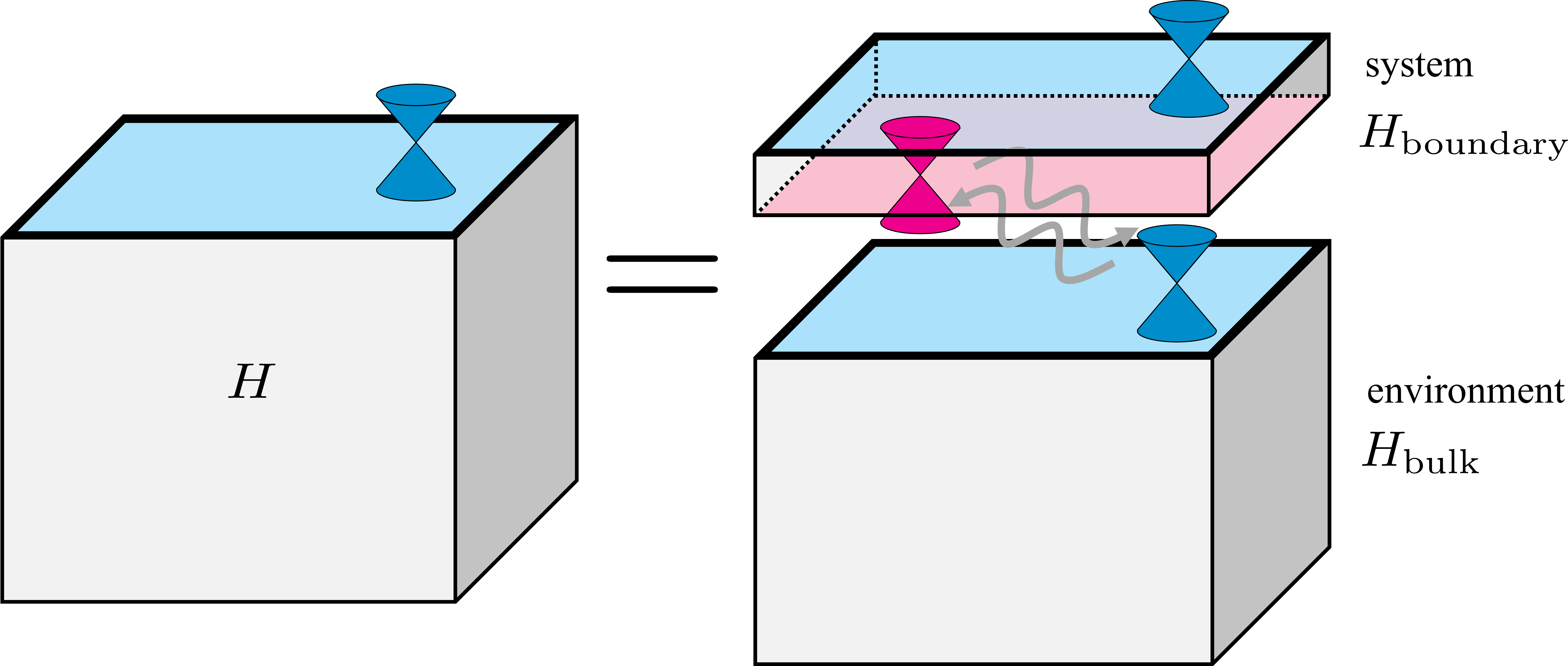}
  \caption{Mixing between a topological state on the bottom boundary of $H_{\rm boundary}$ and a topological state on the top boundary of $H_{\rm bulk}$ gives rise to non-Hermiticity through the self-energy $\Sigma$. The effective boundary Hamiltonian is given by $H_{\rm eff}=H_{\rm boundary}+\Sigma$. 
  \label{fig101-1}}
  \end{center}
\end{figure}

In this section, by regarding the boundary of the Hermitian topological insulator as the system and the bulk as the environment, respectively (Fig.\ref{fig101-1}), we explain how the effective boundary Hamiltonian exhibits non-Hermitian topology.

\subsection{S1.1. Derivation of the boundary effective non-Hermitian Hamiltonian}

In this section, we derive the effective boundary non-Hermitian Hamiltonian.

Consider a system described by a Hermitian Hamiltonian $H$. As shown in Fig.\ref{fig101-1}, we divide the system into the boundary and the environment(= bulk) as follows:
\begin{equation}
    H = \begin{pmatrix}
        H_{\rm boundary} & T^{\dag} \\
        T & H_{\rm bulk}
    \end{pmatrix},
        \label{eq:102-1}
\end{equation}
where $H_{\rm boundary}$ is the boundary part of the Hamiltonian, $H_{\rm bulk}$ is the bulk part of the Hamiltonian, and the matrix $T$ captures the coupling between the bulk and boundary. The (retarded) Green's function $G$ of the total Hermitian Hamiltonian $H$ is defined as
\begin{equation}
    \left[ \left( \omega + \ii \eta \right) - H \right] G = \bm{1},
    \label{eq:102-2}
\end{equation}
where $\eta$ is an infinitesimal positive number reflecting causality, and $\omega$ is the frequency. From Eq. (\ref{eq:102-1}), Eq.(\ref{eq:102-2}) can be rewritten as:
\begin{equation}
\label{eq:102-3}
    \begin{pmatrix}
        \omega + \ii \eta - H_{\rm boundary} & -T^{\dag} \\
        -T & \omega + \ii \eta - H_{\rm bulk}
    \end{pmatrix} \begin{pmatrix}
        G_{\rm boundary} & G_{\rm boundary-bulk} \\
        G_{\rm bulk-boundary} & G_{\rm bulk}
    \end{pmatrix} = \begin{pmatrix}
        \bm{1} & \bm{0} \\
        \bm{0} & \bm{1}
    \end{pmatrix},
\end{equation}
which leads to
\begin{align}
    \left( \omega + \ii \eta - H_{\rm boundary} \right) G_{\rm boundary} - T^{\dag} G_{\rm bulk-boundary} &= \bm{1}, \label{eq:103-1}\\
    -T G_{\rm boundary} + \left( \omega + \ii \eta - H_{\rm bulk} \right) G_{\rm bulk-boundary} &= \bm{0}. \label{eq:103-2}
\end{align}
Since Eq. (\ref{eq:103-2}) implies
\begin{equation}
\label{Gbulkboundary}
    G_{\rm bulk-boundary} 
    = \left( \omega + \ii \eta - H_{\rm bulk} \right)^{-1} T G_{\rm boundary},
\end{equation}
then, substituting Eq.(\ref{Gbulkboundary}) into Eq.(\ref{eq:103-1}), we get
\begin{equation}
\label{boundary-green}
\left( \omega + \ii \eta - H_{\rm boundary}-T^{\dag}(\omega + \ii \eta - H_{\rm bulk})^{-1}T \right)G_{\rm boundary}=\bm{1}.
\end{equation}
The above equation implies that the effective boundary Hamiltonian, defined by
\begin{equation}
    H_{\rm eff}(\omega) \coloneqq H_{\rm boundary}+ T^{\dag} \left( \omega + \ii \eta - H_{\rm bulk} \right)^{-1} T\coloneqq H_{\rm boundary}+\Sigma(\omega)\label{eq:104-1},
\end{equation}
describes the Green's function on the boundary.
Here, we have introduced the self-energy $\Sigma(\omega)$, reflecting effects of the bulk, as
\begin{equation}
\label{eq:104-2}
\Sigma(\omega) \coloneqq T^{\dag} \left( \omega + \ii \eta - H_{\rm bulk} \right)^{-1} T
= T^{\dag} \left( \sum_{n} \frac{\ket{\psi_{n}} \bra{\psi_{n}}}{\omega + \ii \eta - E_{n}}\right) T,
\end{equation}
where $E_n$ and $\ket{\psi_n}$ represent the eigenvalues and eigenstates of $H_{\rm bulk}$, respectively. The self-energy $\Sigma(\omega)$, which indicates the particle exchange between the bulk and boundary, gives non-Hermiticity to the effective boundary Hamiltonian.

If the original Hamiltonian $H$ hosts time-reversal symmetry (TRS), particle-hole symmetry (PHS), and/or chiral symmetry (CS), 
\begin{align}
\mbox{TRS}: {\cal T}H^{*}{\cal T}^{-1}=H, 
\quad \mbox{PHS}: {\cal C}H^t{\cal C}^{-1}=-H, 
\quad \mbox{CS}: \Gamma H \Gamma^{-1}=-H,
\label{eq: AZ}
\end{align}
with unitary operators ${\cal T}$, ${\cal C}$ and $\Gamma$, 
one can prove that the effective Hamiltonian supports TRS$^\dagger$,  
PHS$^\dagger$, and CS defined by
\begin{align}
\mbox{TRS$^\dagger$}: {\cal T}H_{\rm eff}^t(\omega){\cal T}^{-1}=H_{\rm eff}(-\omega), 
\quad \mbox{PHS$^\dagger$}: {\cal C}H_{\rm eff}^*(\omega){\cal C}^{-1}=-H_{\rm eff}(-\omega), 
\quad \mbox{CS}: \Gamma H_{\rm eff}^\dagger(\omega) \Gamma^{-1}
=-H_{\rm eff}(\omega),
\label{eq: AZd}
\end{align}
since the unitary operators ${\cal T}$, ${\cal C}$ and $\Gamma$ act only on the internal degrees of freedom \cite{hamanaka2024nonhermitiantopologyhermitiantopological}. 

To study states near the Fermi energy $E_{\rm F}$, we can approximate $H_{\rm eff}(\omega)$ by $H_{\rm eff}(E_{\rm F})$.
In particular, we choose the origin of the real part of the energy as $E_{\rm F}=0$ and use $H_{\rm eff}\coloneqq H_{\rm eff}(0)$ in our study.

\subsection{S1.2. Hermitian bulk--non-Hermitian boundary correspondence}
In this section, we explain how the topology of the Hermitian Hamiltonian $H$ in $d$ dimensions is related to the topology of the effective boundary Hamiltonian $H_{\rm eff}$ in $(d-1)$ dimensions \cite{inaka-point, hamanaka2024nonhermitiantopologyhermitiantopological}.

The basic idea is as follows: From the bulk-boundary correspondence, when the Hermitian Hamiltonian $H$ describes a $d$-dimensional topological insulator (or a topological superconductor), it hosts a $(d-1)$-dimensional topological gapless boundary state, as illustrated in the left-hand side of Fig.\ref{fig101-1}.
Then, when we divide $H$ into $H_{\rm boundary}$ and $H_{\rm bulk}$, $H_{\rm boundary}$ and $H_{\rm bulk}$ also support the same topological boundary states since they also describe the same topological insulator (superconductor).
See the right-hand side of Fig.\ref{fig101-1}.
If we neglect the coupling $T$ between $H_{\rm boundary}$ and $H_{\rm bulk}$,
the topological states on the top and the bottom boundaries of $H_{\rm boundary}$ become gapful because they are mixed, while the topological boundary states of $H_{\rm bulk}$ remain gapless.

When we turn on $T$ in the effective Hamiltonian, the self-energy $\Sigma(\omega)$ in Eq.(\ref{eq:104-2}) gives a non-Hermitian mixing between the topological boundary states of $H_{\rm boundary}$ and $H_{\rm bulk}$. 
In particular, the bottom (top) boundary state of $H_{\rm boundary}$ and the top boundary state of $H_{\rm bulk}$ are firmly (rarely) mixed since their topological numbers are canceled (the same). 
(More precisely, for a $\mathbb{Z}_2$ topological insulator, both the top and bottom boundary states of $H_{\rm boundary}$ have the same $\mathbb{Z}_2$ number identical to that of the top boundary state of $H_{\rm bulk}$. However, even in this case, only one of the states in $H_{\rm boundary}$ is firmly mixed with the top boundary state of $H_{\rm bulk}$, while the other is not because these states are protected by the $\mathbb{Z}_2$ number.)
Furthermore, 
for $\omega$ near the Fermi energy, the non-Hermitian mixing dominates over others since Eq.(\ref{eq:104-2}) 
is divergent as $-i/\eta$ for gapless states in $H_{\rm bulk}$. 
Thus, we can now neglect the mixing between the top and bottom boundary states of $H_{\rm boundary}$, mentioned above. 
As a result, the top and bottom boundary states in $H_{\rm boundary}$ are back to gapless in the real part of the energy. The bottom one also has the imaginary part of the energy due to the non-Hermitian mixing.
Therefore, the bottom boundary state decays rapidly, and thus, $H_{\rm eff}$ effectively describes only the gapless top boundary state.

This implies that $H_{\rm eff}$ is topologically nontrivial since the top boundary state is topologically nontrivial.
As was argued in Refs.\cite{inaka-point,
hamanaka2024nonhermitiantopologyhermitiantopological}, the corresponding topological number of $H_{\rm eff}$ is given by the point-gap topological number, defined by a gap intrinsic to non-Hermitian Hamiltonians.

To illustrate the above argument, let us consider the Qi-Wu-Zhang model for a Chern insulator \cite{QWZ, TQFT-2008-TRS}:  
\begin{align}
\label{chern}
    H_{\mathrm{Chern}} \left( \bm{k} \right) = \left( t \sin{k_x} \right) \sigma_x + \left( t \sin{k_y} \right) \sigma_y \nonumber + \left( m + t \cos{k_x}+t \cos{k_y} \right) \sigma_z,
\end{align}
where $\sigma_i$'s ($i=x, y, z$) are the Pauli matrices, and $t,m \in \mathbb{R}$. 
The first Chern number $C_1$ of the model is $C_1={\rm sgn} \left(m/t\right)$ for $\left|m/t\right|<2$, and $C_1=0$ for $\left|m/t\right|>2$. 
Below, we consider the $C_1=1$ case. 

We first divide the Hermitian system into the boundary and the bulk. 
For this purpose, we introduce the lattice representation of the system and  
treat the $x = 0$ part of the Hamiltonian as the boundary and the remaining $x \ge 1$ part as the bulk. 
Here, we take the convention where $x=0$ is the top of the system, and $x$ increases when going down.
The boundary Hamiltonian is a single layer in this case, so its top and bottom boundaries are identical. Nonetheless, our argument above works, as shown below.

Using Eq. (\ref{eq:104-1}), we derive the effective boundary Hamiltonian $H_{\rm eff}$ and compute the energy spectrum.
The blue curve in Fig.\ref{fig101-1} (a) shows the complex spectrum of $H_{\rm eff}$ under the periodic boundary condition in the $y$-direction ($y$-PBC). 
The curve does not have a gap in the real part of the energy, and it crosses the ${\rm Re}E=0$ line twice, corresponding to the top and bottom boundary gapless states of $H_{\rm eff}$:
One of the crossing points between the blue curve and the ${\rm Re}=0$ line is real, corresponding to the top boundary state, and the other has a negative imaginary part, corresponding to the bottom boundary state. 
Moreover, the blue curve shows a loop in the complex energy plane. The spectral winding number of the curve 
\begin{equation}
    W= 
    \int_{-\pi}^{\pi} \frac{dk_y}{2\pi\ii} \left( \frac{\partial}{\partial k_y} \log \det \left[ H \left( k_y \right) - E_0 \right] \right),
\end{equation}
is the point-gap topological number of $H_{\rm eff}$, and $W=1$ for $E_0$ inside the loop. 
The spectral winding number $W$ coincides with the Chern number $C_1$ since the winding number is determined by the number of the top boundary states. 
These results are consistent with our argument above.

\begin{figure}[h]
\begin{center}
\includegraphics[scale=0.15]{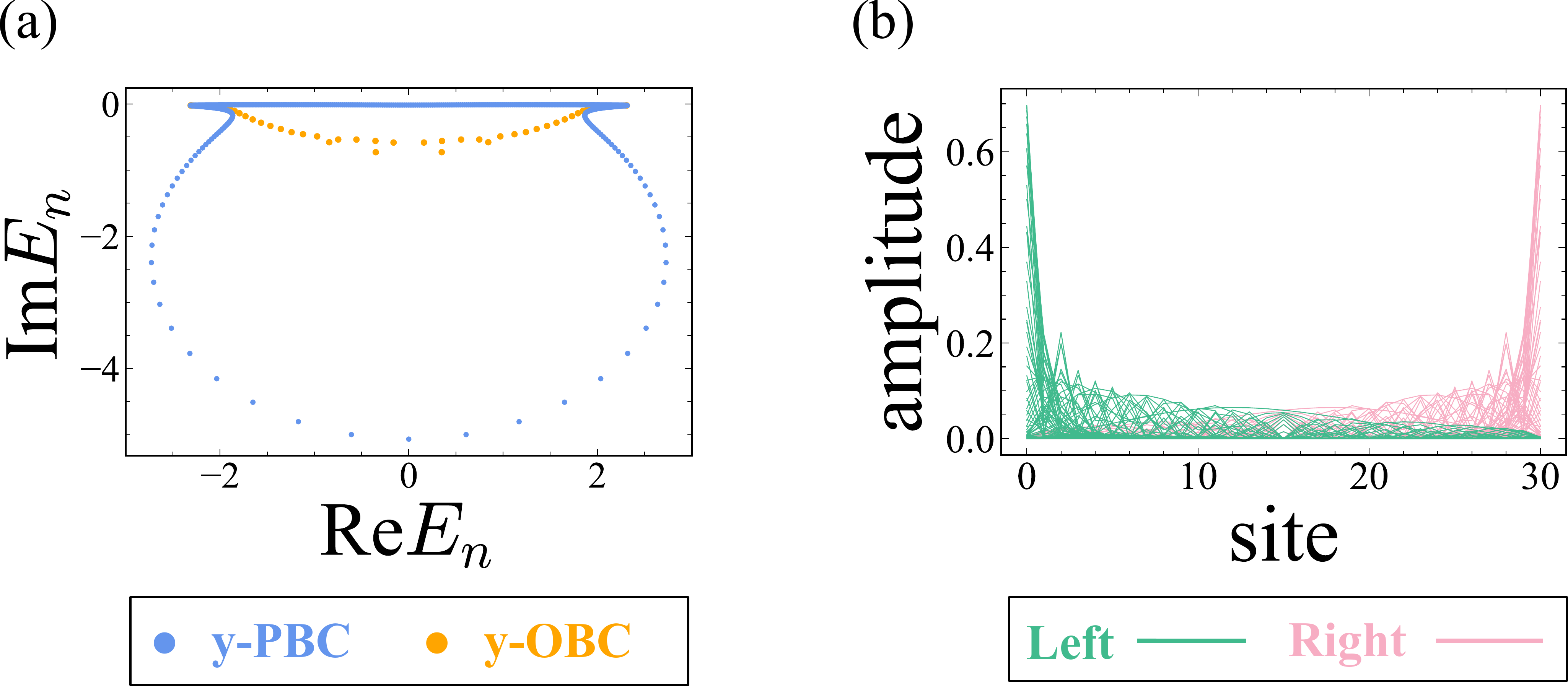}
\caption{Complex energy spectrum of the effective boundary Hamiltonian and the corresponding eigenstates for the Chern insulator ($t = 1.0, m = -1.3, L_x=31$). (a) Complex energy spectrum of $H_{\rm eff}$ under the $y$-PBC ($L_y = 300$) and $y$-OBC ($L_y = 31$). 
(b) Left and right eigenstates of $H_{\rm eff}$ under the $y$-OBC. The infinitesimal parameter $\eta$ is chosen as $\eta = 1/\sqrt{L_x}$ \cite{hamanaka2024nonhermitiantopologyhermitiantopological}. \label{fig106-1}}
\end{center}
\end{figure}

The nonzero spectral winding number of $H_{\rm eff}$ results in the extreme sensitivity of the spectrum against the boundary condition \cite{okss,Zhang-2020}
, called the non-Hermitian skin effect \cite{lee-skin-2016, yao-skin-2018}.
The orange arc in Fig. \ref{fig106-1} (a) is the complex spectrum of $H_{\rm eff}$ under the open boundary condition in the $y$-direction ($y$-OBC), which is entirely different from the spectrum under the $y$-PBC (blue curve).
The left and right eigenstates of $H_{\rm eff}$ under the $y$-OBC are localized at opposite endpoints in the $y$-direction, 
as shown in Fig.\ref{fig106-1}.
These unique localization modes provide a positive or negative current flow between the boundary and the bulk at corners, consistent with the edge current by the chiral edge state of the Chern insulator \cite{hamanaka2024nonhermitiantopologyhermitiantopological}.

\section{S2. Callan-Rubakov effects for monopoles}
The Callan-Rubakov effect is a catalysis of proton decay, which originates from a strange property of a $J=0$ chiral fermion (namely Weyl fermion) coupled to a monopole. 
Here, we briefly review this strange property. 
Whereas the original Callan-Rubakov effect considers the 't Hooft-Polyakov monopole in the SU(5) grand unified theory, the strange property is already evident in the Dirac monopole in the U(1) electro-magnetic gauge theory \cite{Kazama1977}. Thus, for simplicity, we consider only the latter case here. 

Let us consider the Dirac monopole given by
\begin{align}
    \bm{B}=g\frac{\bm{r}}{r^{3}},
\end{align}
where ${\bm r}=(x,y,z)$ is the three-dimensional coordinate, and $r=|{\bm r}|$ is the magnitude of ${\bm r}$.
From the Dirac quantization, the monopole charge $g$ satisfies
\begin{align}
    ge=q \in \frac{1}{2}\mathbb{Z},
\end{align}
where $e$ is the electric charge of the Weyl fermion we consider.
In the presence of the monopole, the angular momentum $\bm{L}$ is given by
\begin{align}
    \bm{L}=\bm{r}\times(\bm{p}-e\bm{A})-q\bm{r}/r\equiv (L_x,L_y,L_z),\quad \bm{p}=-i\hbar\bm{\nabla},
\end{align}
which satisfies the conventional commutation relations for the angular momentum, 
\begin{align}
[L_i,L_j]=i\epsilon_{ijk}L_k,    
\end{align}
where we have used the Einstein summation convention, and $\epsilon_{ijk}$ is the Levi-Civita symbol. 
Thus, for spin 1/2 fermions coupled to the monopole, the total angular momentum ${\bm J}$ is given by
\begin{align}
    \bm{J}=\bm{r}\times(\bm{p}-e\bm{A})-q\bm{r}/r+{\bm s}/2,    
\end{align}
where ${\bm s}=(s_x,s_y,s_z)$ is the vector of the Pauli matrices in the spin space. 
Below, we consider a monopole with the minimal charge, $g=1/2|e|$, where ${\bm J}$ is given by
\begin{align}
    \bm{J}=\bm{r}\times(\bm{p}-e\bm{A})-\frac{{\rm sgn}[e]}{2}\frac{\bm{r}}{r}
    +\frac{{\bm s}}{2}.        
\label{eq: J}
\end{align}
${\bm J}^2$ is quantized as $J(J+1)$, but $J$ is an integer, not a half-integer, due to the influence of the monopole.

We can see the oddness of the $J=0$ Weyl fermion immediately from Eq. (\ref{eq: J}). For $J=0$, we have ${\bm J}\cdot{\bm r}=0$, which implies that the $J=0$ fermion obeys
\begin{align}
{\rm sgn}[e]={\bm s}\cdot {\bm r}/r.        
\end{align}
Thus, for a $J=0$ Weyl fermion incoming to the monopole,  which satisfies ${\bm r}/r=-{\bm p}/p$ with ${\bm p}$ and $p$ the momentum of the fermion and its magnitude, we have 
\begin{align}
{\rm sgn}[e]=-{\bm s}\cdot {\bm p}/p.
\end{align}
Therefore, if the incoming Weyl fermion has a positive (negative) charge $e>0$ ($e<0$), its helicity ${\bm s}\cdot {\bm p}/p$ must be negative (positive).  
Similarly, for a $J=0$ Weyl fermion outgoing from the monopole, we have 
\begin{align}
{\rm sgn}[e]={\bm s}\cdot {\bm p}/p,        
\end{align}
which implies that if the outgoing Weyl fermion has a positive (negative) charge $e>0$ ($e<0$), its helicity ${\bm s}\cdot {\bm p}/p$ must be positive (negative).
Because the helicity (positive or negative) and the chirality (right-handed or left-handed) coincide for Weyl fermions, only particular combinations between the charge and the chirality are allowed for the $J=0$ incoming and outgoing Weyl fermions, as summarized in Table \ref{tab: Weyl+monopole}: 
The $J=0$ left-handed (right-handed) Weyl fermion
with charge +1 (-1) is a purely incoming wave,
while the $J=0$ right-handed (left-handed) one with charge +1 (-1)
is a strictly outgoing wave.
In particular, the outgoing Weyl fermion has different combinations of charge and chirality from those of the incoming Weyl fermion. 
Therefore, once the $J=0$ Weyl fermion is scattered by the monopole, it must change its charge or chirality. 
The same constraint exists for the chiral fermions coupled to the 't Hooft-Polyakov monopole in the grand unified theory, which gives rise to the proton-decay catalysis. (For a review, see Ref.\cite{Rubakov-review-88}.)

\begin{table*}[t]
	\centering
	\caption{$J=0$ Weyl fermion coupled to monopole.} 
	\label{tab: Weyl+monopole}
     \begin{tabular}{c|c|c} \hline \hline
    Direction of motion & Electric charge & Chirality (helicity)\\ 
     \hline
   incoming & +1 & left-handed (-)\\
   incoming & -1 & right-handed (+)\\ 
   outgoing & +1 & right-handed (+)\\
   outgoing & -1 & left-handed (-)\\
   \hline \hline
  \end{tabular}
\end{table*}

\section{S3. Details of numerical calculations in Figs. \ref{fig_flux_hermitian_non-Hermitian_eigenstate_LDOS} and \ref{fig_edge_screw_energy_eigenstate_LDOS}}

\subsection{S3.1. Line defect configurations}
Here, we describe the details of the line defects in the calculations of Figs. \ref{fig_flux_hermitian_non-Hermitian_eigenstate_LDOS} and \ref{fig_edge_screw_energy_eigenstate_LDOS}. 

\begin{figure}[btp]
 \begin{center}
  \includegraphics[scale=0.0630]{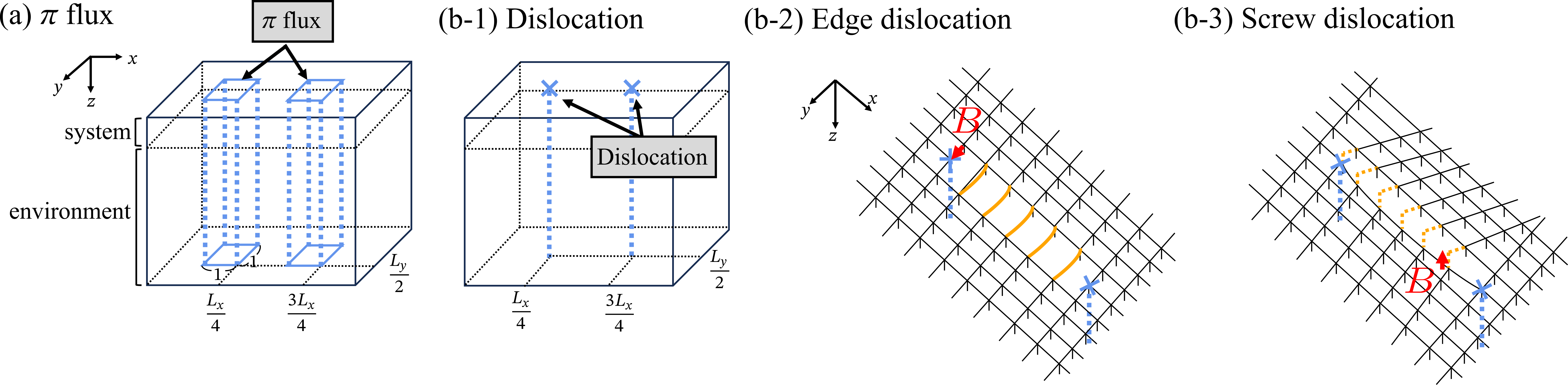}
  \caption{Schematic illustrations of the line defects in the topological insulator (\ref{eq:hamil-TI}). (a) $\pi$-flux: Two $\pi$-fluxes are introduced at the positions indicated in the figure along the $z$-direction. To ensure the PBC in the $x$ and $y$ directions, we consider two $\pi$-fluxes. (b-1) Dislocation: The blue dots represent the positions of the edge or screw dislocations. Both dislocations are introduced for each layer $z$. (b-2) Edge dislocation: Blue cross marks correspond to the positions marked in Fig. (b-1). $\bm{B}$ represents the Burgers vector ($\bm{B}=\hat{\bm y}$). (b-3) Screw dislocation: Blue cross marks correspond to the positions marked in Eq. (b-1). $\bm{B}$ represents the Burgers vector ($\bm{B}=\hat{\bm z}$). \label{sketch_defect}}
  \end{center}
\end{figure}

For Fig.\ref{fig_flux_hermitian_non-Hermitian_eigenstate_LDOS}, we consider two $\pi$-fluxes oriented along the $z$ direction at the positions indicated in Fig. \ref{sketch_defect} (a). 
We insert two $\pi$-fluxes to ensure the periodic boundary conditions in the $x$ and $y$ directions.
The numerical implementation of $\pi$-fluxes follows the method in Ref. \cite{Schindler-magnetic2023}. 

For Fig.\ref{fig_edge_screw_energy_eigenstate_LDOS}, we introduce dislocations at the positions depicted in Fig. \ref{sketch_defect} (b-1). 
The Burgers vectors for the edge dislocations and the screw dislocations are shown in \ref{sketch_defect} (b-2) and (b-3), respectively.
The position of blue dots in Fig. \ref{sketch_defect} (b-1) is the same as that in Fig. \ref{sketch_defect} (b-2) and (b-3).
The numerical implementation of these dislocations follows the methods in Refs. \cite{schindler2022defectNature,imura-weak2011,ran2009dislocation}.

\subsection{S3.2. Line-defect skin modes}
\begin{figure}[btp]
 \begin{center}
  \includegraphics[scale=0.058]{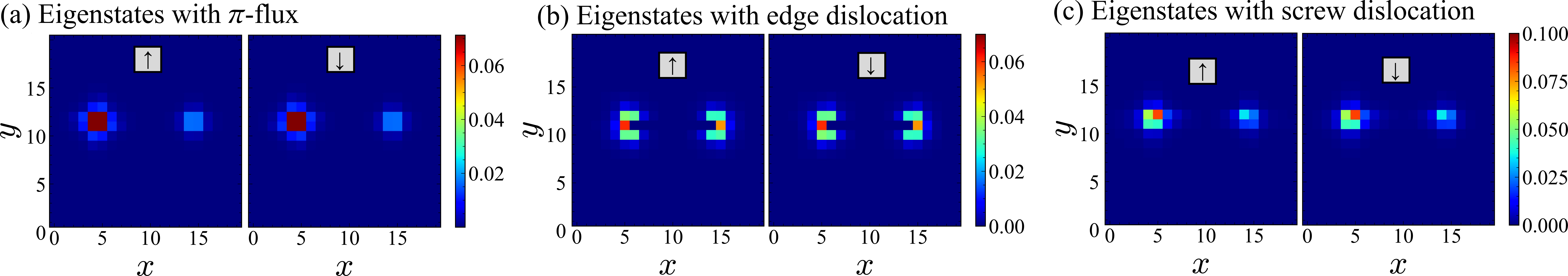}
  \caption{Amplitude of the right eigenstates ($L_x=L_y=20, L_z=10, t=1.0, \omega=0$). Each eigenstate is plotted separately for each spin degrees of freedom. The periodic boundary conditions are imposed along the $x$ and $y$ directions, while the open boundary condition is applied along the $z$ direction. The infinitesimal number $\eta$ is chosen as $\eta=1/\sqrt{L_z}$ \cite{hamanaka2024nonhermitiantopologyhermitiantopological}. (a) $\pi$-flux skin modes ($m=-2.0$). The corresponding eigenvalue is $E_n = -0.31 - 0.57\ii$. (b) Edge dislocation skin modes ($m=2.0$). The corresponding eigenvalue is $E_n=0.74-0.56 \ii$. (c) Screw dislocation skin modes ($m=2.0$). The corresponding eigenvalue is $E_n=-1.37 \ii$.
      \label{eigenstates}}
 \end{center}
\end{figure}

Figure \ref{eigenstates} shows the right eigenstates associated with the skin modes of $H_{\rm eff}$ in the presence of line defects. 
Figures \ref{eigenstates} (a), (b), and (c) are typical profiles of the skin modes on the right-hand side of Fig. 
\ref{fig_flux_hermitian_non-Hermitian_eigenstate_LDOS}(b), the left-hand side of Fig. \ref{fig_edge_screw_energy_eigenstate_LDOS} (a), and the right-hand side of 
Fig. \ref{fig_edge_screw_energy_eigenstate_LDOS} (a), respectively.
Due to the skin effect, these eigenstates are localized around the line defects.

\subsection{S3.3. Details on local density of state in the presence of $\pi$-flux}
As discussed in the main text, the surface mode relevant to the Callan-Rubakov effect in topological insulators has the enhanced local density of states at the intersection between a line defect and the surface. 
In particular, for the $\pi$-flux, we have argued that the enhanced local density of states 
shows the power-law behavior $1/r$ as a function of the distance $r$ from the $\pi$-flux (see below Eq. (\ref{eq:hamil-TI})). 
Here, we numerically confirm this behavior.

\begin{figure}[btp]
 \begin{center}
  \includegraphics[scale=0.088]{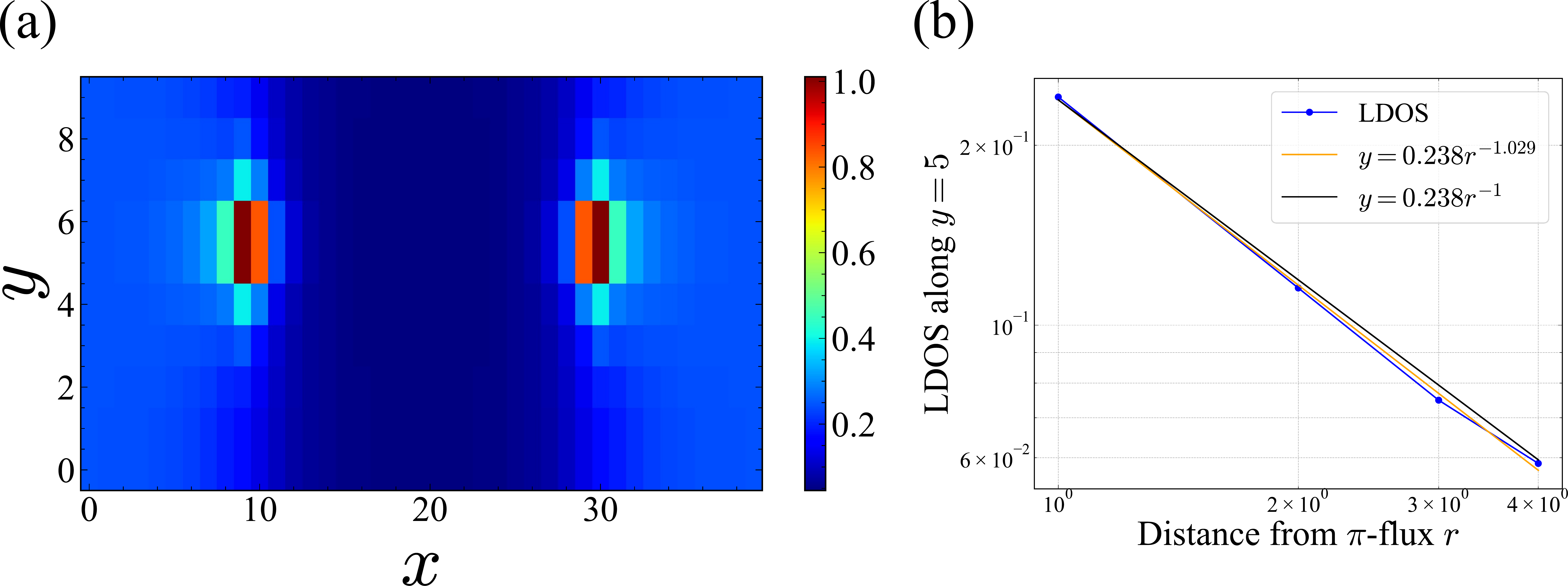}
  \caption{Local density of states on the surface of the model of a three-dimensional topological insulator (\ref{eq:hamil-TI}) with $\pi$-flux ($L_x=40, L_y=10, L_z=10, t=1.0, m=-2.0, \omega=0$). The periodic boundary conditions are imposed along the $x$ and $y$ directions, while the open boundary condition is applied along the $z$ direction. The blue dots in (b) represent the numerical results of the local density of states. The orange line is the best power-law fitting.
      \label{fig_LDOSrDep}}
 \end{center}
\end{figure}
Figure \ref{fig_LDOSrDep} indicates the local density of states for the model (\ref{eq:hamil-TI}) with the parameters, $L_x=40, L_y=10, L_z=10, m=-2, \omega=0$. 
We choose $\eta$ in $H_{\rm eff}$ of Eq.(\ref{eq:104-1}) as $\eta=0.11$, satisfying the condition $1/L_z<\eta<\mathcal{O}(L_z^0)$ \cite{hamanaka2024nonhermitiantopologyhermitiantopological, Datta-textbook-1995, Datta-textbook-2005}, then, use the retarded Green's function on the boundary,
$G_{\rm boundary}=(\omega-H_{\rm eff})^{-1}$, to calculate the local density of states.
Like the case in the main text, we find the enhanced local density of states
around the $\pi$-flux (Fig. \ref{fig_LDOSrDep}(a)).
Figure \ref{fig_LDOSrDep}(b) is the local density of states along $y=5$ as a function of the distance $r$ from the $\pi$-flux, consistent with the $1/r$ behavior.

\section{S4. dislocation modes and dislocation skin effects}
In the main text, we have shown numerically that if a dislocation supports a one-dimensional helical mode, a surface normal to the dislocation exhibits a dislocation non-Hermitian skin effect. 
Here, we analytically prove this claim in the case of edge dislocations by relating the condition for dislocation modes \cite{ran2009dislocation} with the condition for dislocation skin effects \cite{Schindler-dislocation2021}.

The condition for dislocation modes in three-dimensional topological insulators is as follows. 
Consider a three-dimensional topological insulator with the strong index $\nu_0$ and the weak indices ${\bm M}_\nu=(\nu_x,\nu_y,\nu_z)$ and insert a dislocation with the Burgers
vector ${\bm B}$. 
Then, Ran {\it et al.} have shown that the dislocation supports a one-dimensional helical mode if the following condition is satisfied \cite{ran2009dislocation}:
\begin{align}
{\bm B} \cdot {\bm M}_\nu = \pi \,\mbox{(mod. $2\pi$)}.     
\label{eq: ran}
\end{align}

Meanwhile, Schindler and Perm have discussed the condition for dislocation skin effects in two-dimensional class AII$^\dagger$ point-gapped non-Hermitian systems \cite{Schindler-dislocation2021}. 
The effective Hamiltonian on the surface of a three-dimensional time-reversal invariant topological insulator has the transpose version of time-reversal symmetry \cite{inaka-point, hamanaka2024nonhermitiantopologyhermitiantopological}. Thus, it belongs to class AII$^\dagger$ \cite{KSUS}. 
Although the effective Hamiltonian is gapless in the real part of the energy since it describes gapless surface Dirac fermions, it has a non-Hermitian gap in the complex energy plane called a point gap \cite{KSUS}. 
Therefore, one can apply the condition in Ref.\cite{Schindler-dislocation2021} to the effective Hamiltonian on the surface of a three-dimensional time-reversal invariant topological insulator.
In Table \ref{tab: class AIId}, we show the topological table for class AII$^\dagger$ point gapped non-Hermitian systems.
From this table, we find that a two-dimensional class AII$^\dagger$ point-gapped Hamiltonian has 
a two-dimensional strong $\mathbb{Z}_2$ index $\nu_0^{\rm NH}$ and two weak $\mathbb{Z}_2$ indices
${\bm M}_\nu^{\rm NH}=(\nu_x^{\rm NH}, \nu_y^{\rm NH})$.
Here, the weak $\mathbb{Z}_2$ index $\nu_{a=x,y}^{\rm NH}$ 
is given by the $d=1$ $\mathbb{Z}_2$ index in Table \ref{tab: class AIId} on the $k_{a=x,y}=\pi$ line in the surface Brillouin zone.
Following the argument by Schindler and Perm, we can show that if the two-dimensional system has an edge dislocation with two-dimensional Burgers vector ${\bm B}_{\rm 2d}$, it shows the dislocation skin effect if the following condition is met: 
\begin{align}
{\bm B}_{\rm 2d}\cdot{\bm M}_\nu^{\rm NH}=\pi \,\mbox{(mod. $2\pi$)}. 
\label{eq: schindler}
\end{align}
Here, we have used a different notation for the weak indices: 
$v_x(E,\pi)$ in Ref.\cite{Schindler-dislocation2021} coincides with $e^{i\nu_y^{\rm NH}}$ in our notation, and we obtain Eq.(\ref{eq: schindler}) by generalizing Eq. (7) in Ref. \cite{Schindler-dislocation2021} to our case. 

\begin{table*}[t]
	\centering
	\caption{Topological table for class AII$^\dagger$ point-gapped non-Hermitian systems.} 
	\label{tab: class AIId}
     \begin{tabular}{c|ccc} \hline \hline
   spatial dimension  &$d=0$ & $d=1$ & $d=2$\\ 
     \hline
   topological index &0 & $\mathbb{Z}_2$ & $\mathbb{Z}_2$\\
   \hline \hline
  \end{tabular}
\end{table*}

\begin{figure}[btp]
 \begin{center}
  \includegraphics[scale=0.1]{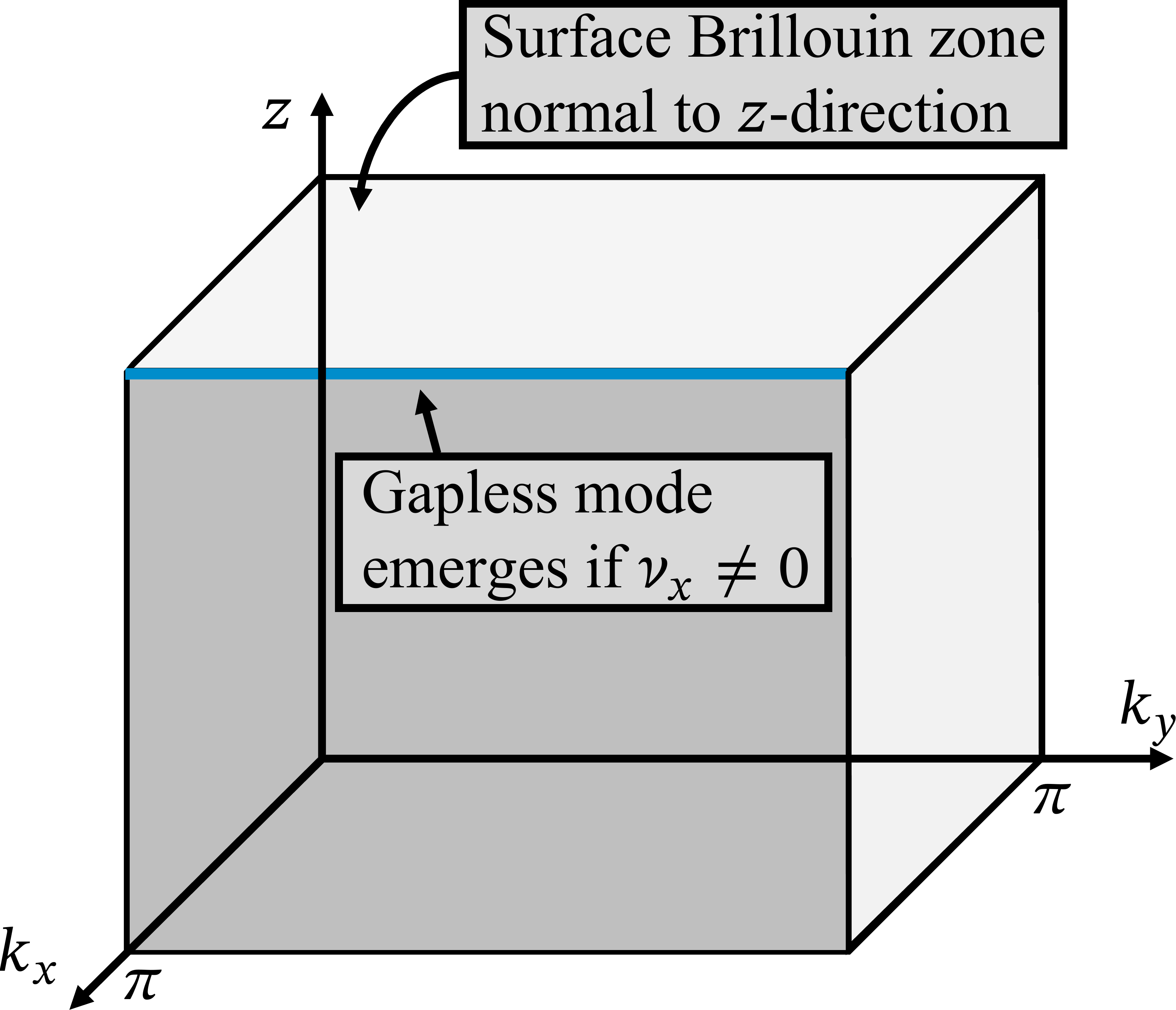}
  \caption{The weak index $\nu_x$ in a three-dimensional time-reversal invariant topological insulator and the corresponding topological boundary state.
      \label{fig_dis_modes_skin_modes}}
 \end{center}
\end{figure}

Now we relate Eq.(\ref{eq: ran}) 
with Eq.(\ref{eq: schindler}). 
Since Eq.(\ref{eq: schindler}) does not apply to screw dislocations (because a purely two-dimensional system does not have screw dislocations), we focus on Eq.(\ref{eq: ran}) with edge dislocations. 
We also assume that the edge dislocation goes along the $z$-direction and consider a surface normal to the dislocation.
Thus, the two-dimensional momentum $(k_x,k_y)$ describes the surface normal to the dislocation, and the Burgers vector ${\bm B}$ has the form
${\bm B}=({\bm B}_{\rm 2d},0)$.
In this case, the condition in Eq.(\ref{eq: ran}) reduces to
\begin{align}
{\bm B}_{\rm 2d}\cdot (\nu_x,\nu_y)=\pi\,(\mbox{mod. $2\pi$}),     
\end{align}
which coincides with Eq.(\ref{eq: schindler}) if $\nu_a=\nu_a^{\rm NH}$ $(a=x,y)$.
The last equation $\nu_a=\nu_a^{\rm NH}$ becomes evident once one considers the meaning of these topological indices. We first prove $\nu_x=\nu_x^{\rm NH}$ for concreteness. 
Since the weak index $\nu_x$ for a three-dimensional time-reversal invariant topological insulator is given by the $\mathbb{Z}_2$ index defined on the two-dimensional Brillouin zone with $k_x=\pi$, the surface of the topological insulator hosts a gapless mode along the $k_x=\pi$ line if $\nu_x$ is nontrivial. See Fig. \ref{fig_dis_modes_skin_modes}.  
Therefore, if $\nu_x$ is non-trivial, 
the effective Hamiltonian on the surface is also topologically nontrivial on the $k_x=\pi$ line on the surface Brillouin zone. Thus, we have $\nu_x=\nu_x^{\rm NH}$. We can prove $\nu_y=\nu_y^{\rm NH}$ similarly.

\section{S5. Bosonization analysis of Eq.(\ref{eq:Schwinger})}
Here, we examine the variant of the massless Schwinger model in Eq.(\ref{eq:Schwinger})
\begin{align}
L_{J_z=0}=\int dr\left[-\frac{\pi r}{2e^2}f_{\alpha\beta}f^{\alpha\beta}+i\bar{\chi}\gamma^\alpha
(\partial_\alpha-ia_\alpha)\chi\right],   
\end{align}
by using bosonization. 
The standard bosonization dictionary reads (see, for example, Ref.\cite{fradkin2010field})
\begin{align}
i\bar{\chi}\gamma^\alpha\partial_\alpha\chi 
\Leftrightarrow
\frac{1}{8\pi}\partial^\alpha\phi\partial_\alpha\phi,
\quad
-\bar{\chi}\gamma^\alpha\chi
\Leftrightarrow -\frac{1}{2\pi}\epsilon^{\alpha\beta}\partial_\beta\phi,
\quad
\bar{\chi}\chi
\Leftrightarrow -\frac{1}{\pi\epsilon}\cos\phi,
\end{align}
where $\phi$ is a real bosonic field and $\epsilon$ is a UV length-scale.  Thus, we can rewrite the above model as
\begin{align}
L_{J_z=0}&=\int dr \left[-\frac{\pi r}{2e^2}f_{\alpha\beta}f^{\alpha\beta}+\frac{1}{8\pi}\partial^\alpha
\phi\partial_\alpha\phi
+\frac{1}{2\pi}\epsilon^{\alpha\beta}a_\alpha\partial_\beta\phi
\right]
\nonumber\\
&=\int dr \left[-\frac{\pi r}{2e^2}f_{\alpha\beta}f^{\alpha\beta}+\frac{1}{8\pi}\partial^\alpha
\phi\partial_\alpha\phi
+\frac{1}{2\pi}\epsilon^{\alpha\beta}\partial_\alpha a_\beta\phi
\right]
\nonumber\\
&=\int dr \left[-\frac{\pi r}{e^2}f_{tr}^2+\frac{1}{8\pi}\partial^\alpha
\phi\partial_\alpha\phi
+\frac{1}{2\pi}f_{tr}\phi
\right]
\nonumber\\
&=\int dr \left[-\frac{\pi r}{e^2}
\left(f_{tr}-\frac{e^2}{4\pi^2r}\phi
\right)^2
+\frac{1}{8\pi}\partial^\alpha
\phi\partial_\alpha\phi
+\frac{e^2}{16\pi^3r}\phi^2
\right],
\end{align}
where we have performed the partial integral in the second equality and neglected the boundary term.
Then, after integrating out the first term in the last line, the system reduces to the free boson theory with the mass $e^2/(2\pi^2 r)$.
Since the mass diverges at $r=0$, $\langle \phi^n \rangle $ ($n=1,2,3,\dots$) vanishes at the location of the $\pi$-flux.
Therefore, the condensation $|\langle\bar{\chi}\chi\rangle|$ reaches the maximum at the origin:
\begin{align}
|\langle \bar{\chi}\chi \rangle|=\frac{1}{\pi \epsilon}|\langle \cos\phi \rangle|
\xrightarrow{r\to 0}\frac{1}{\pi \epsilon},
\end{align}
where we have used the bosonization again. The nonzero condensation $\langle \bar{\chi}\chi\rangle=\langle \chi^\dagger_+\chi_-+\chi^\dagger_-\chi_+\rangle$ enables the scattering between $\chi_+$ and its Kramers partner $\chi_-$ at the $\pi$-flux.

\section{S6. Callan-Rubakov effect in three-dimensional topological insulators with line defects}
Here, we explain the condition for the Callan-Rubakov effects in three-dimensional topological insulators/superconductors with line defects.
As discussed in the main text, 
such Callan-Rubakov effects need gapless surface fermions and defect skin modes on the surfaces simultaneously. 
Correspondingly, they require the following topological numbers on the bulk Hermitian Hamiltonian $H$ and the surface effective Hamiltonian $H_{\rm eff}$, respectively: (i)
A three-dimensional topological number of
the bulk Hermitian Hamiltonian $H$, which ensures gapless surface fermions, and (ii) a one-dimensional defect
topological number of the surface non-Hermitian Hamiltonian
$H_{\rm eff}$, which is indispensable for defect skin effects. 
Respecting TRS, PHS, and CS in Eq.(\ref{eq: AZ}), 
we can specify the first topological number from the $d=3$ part of the topological table \cite{classification-andreas2008, kitaev2009periodic, teo-kane-defect} for the Altland-Zirnbauer (AZ) classes. (Here, $d=3$ represents the spatial dimension of $H$.)
Furthermore, since $H_{\rm eff}$ hosts TRS$^\dagger$, PHS$^\dagger$, and/or CS in Eq.(\ref{eq: AZd}) if $H$ has TRS, PHS, and/or CS, and the skin effect originates from one-dimensional point-gap topological phases \cite{okss}, we can identify the second topological number from the topological table for point-gap topological phases in AZ$^\dagger$ classes \cite{KSUS}.
(See the $\delta=1$ part of Table I of Ref.\cite{inaka-point}. )
Note that 
we have used the notation $\delta_{\rm eff}=d_{\rm eff}-D$, instead of $\delta$, to distinguish it from $\delta=d-D$ for $H$ in the main text. 
Combining them, we obtain Table \ref{tab: classification} in the main text. 

\begin{figure}[btp]
 \begin{center}
  \includegraphics[scale=0.11]{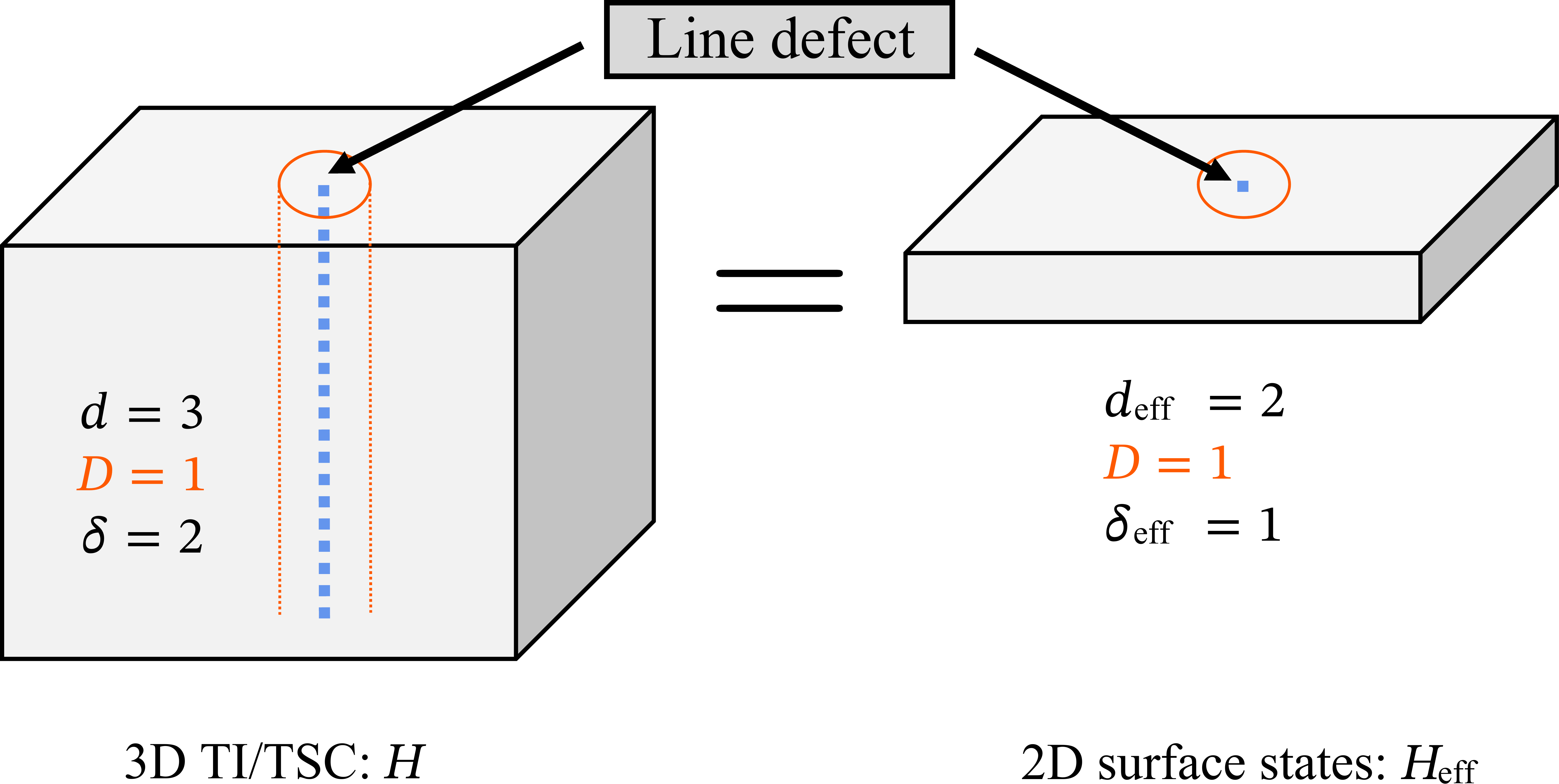}
  \caption{Relation between three-dimensional topological insulators(TIs)/superconductors(TSCs) described by the Hermitian Hamiltonian $H$ and two-dimensional surface modes described by the effective non-Hermitian Hamiltonian $H_{\rm eff}$. 
  The spatial dimension $d_{\rm eff}$ of $H_{\rm eff}$ is $d_{\rm eff}=d-1=2$.
  For a surface normal to the line defect, the codimension $D$ of the defect is $D=1$ both in $H$ and $H_{\rm eff}$. Here, $\delta$ and $\delta_{\rm eff}$ are defined by $\delta=d-D$ and $\delta_{\rm eff}=d_{\rm eff}-D$. 
      \label{fig_TI_with_line}}
 \end{center}
\end{figure}

As explained in 
Sec.S1,
there is a one-to-one correspondence between the Hermitian topology in $H$ and the non-Hermitian one in $H_{\rm eff}$; namely, the right-hand side in Fig.\ref{fig_TI_with_line} is topologically non-trivial if and only if the left-hand side in Fig.\ref{fig_TI_with_line} is topologically non-trivial.
Therefore, a non-trivial $d=3$ topology of $H$ results in a non-trivial $d_{\rm eff}=2$ topology of $H_{\rm eff}$, and  
a non-trivial $\delta_{\rm eff}=1$ topology of $H_{\rm eff}$ results in a non-trivial $\delta=1$ topology of $H$. 
In particular, the latter relation implies that the line-defect skin modes on the surface always accompany gapless modes along the line-defect due to the $\delta=1$ topology.
This result is consistent with the absorption process in Fig. \ref{two possibility} in the main text.

\end{document}